\documentclass[11pt,a4paper]{article}
\usepackage{jheppub}

\usepackage{lineno}
\linenumbers

\usepackage{float} 
\usepackage{subfig} 
\usepackage{amsmath} %

\def\be{\begin{equation}}
\def\ee{\end{equation}}
\def\bea{\begin{eqnarray}}
\def\eea{\end{eqnarray}}
\def\NO{\nonumber}

\def\dfrac{\displaystyle\frac}

\def\b{\beta}

\def\e{\epsilon}

\def\g{\gamma}

\def\s{\sigma}

\newcommand{\beq}{\begin{eqnarray}}
\newcommand{\eeq}{\end{eqnarray}}

\newcommand{\ep}{\varepsilon}

\newcommand{\eqs}[1]{\begin{equation} \begin{split} #1\end{split} \end{equation} }

\newcommand{\ie}{{\it i.e.}}
\newcommand{\eg}{{\it e.g.}}
\newcommand{\etal}{{\it et al.}}
\newcommand{\Q}{{\cal Q}}

\newcommand{\ce}[1]{Eq.~\eqref{#1}}

\newcommand{\cf}[1]{{Fig.~\ref{#1}}}

\def\lsim{\raise0.3ex\hbox{$<$\kern-0.75em\raise-1.1ex\hbox{$\sim$}}}
\def\gsim{\raise0.3ex\hbox{$>$\kern-0.75em\raise-1.1ex\hbox{$\sim$}}}

\def\beq     {\begin{equation}}
\def\eeq     {\end{equation}}

\title{Next-to-leading-order QCD corrections to the yields and polarisations of {\boldmath $J/\psi$} and {\boldmath $\Upsilon$}  
directly produced in association with a {\boldmath $Z$} boson at the LHC}

\author[a,b]{Bin Gong,}
\author[c]{Jean-Philippe Lansberg,}
\author[c,d]{C\'edric Lorc\'e,}
\author[a,b]{Jianxiong Wang}

\affiliation[a]{Institute of High Energy Physics, CAS, P.O. Box 918(4), Beijing, 100049, China}
\affiliation[b]{Theoretical Physics Center for Science Facilities, CAS, Beijing, 100049, China}
\affiliation[c]{IPNO, Universit\'e Paris-Sud, CNRS/IN2P3,
91406, Orsay France}

\affiliation[d]{LPT , Universit\'e Paris-Sud, CNRS,
91406, Orsay France}

\emailAdd{twain@ihep.ac.cn}
\emailAdd{lansberg@in2p3.fr}
\emailAdd{lorce@ipno.in2p3.fr}
\emailAdd{jxwang@ihep.ac.cn}

\abstract{We update the study of the production of direct $J/\psi$ in association with a $Z$ boson at
the Next-to-Leading Order (NLO) in $\alpha_s$ by evaluating both the yield differential
in $P_T$ and the $J/\psi$ polarisation in the QCD-based Colour-Singlet Model (CSM). 
Contrary to an earlier claim, QCD corrections at small and mid $P_T$ are small if one assumes that the factorisation and the
renormalisation scales are commensurate with the $Z$ boson mass. As it can be anticipated, the $t$-channel 
gluon-exchange ($t-$CGE) topologies start to be dominant only for $P_T\, \gsim\, m_Z/2$.  
The polarisation pattern is not altered by the QCD corrections. This is thus far the first quarkonium-production 
process where this is observed in the CSM.  Along the same lines, our predictions for direct $\Upsilon+Z$ are also given.
}

\keywords{$J/\psi$ and $\Upsilon$ production, $Z$ boson, QCD corrections}

\begin{document} 

\maketitle

\section{Introduction}

A few years ago, non-perturbative effects associated with 
colour-octet (CO) channels~\cite{Kramer:2001hh,Brambilla:2004wf,Lansberg:2006dh}
were considered to be the only plausible explanation for the numerous puzzles in the predictions of quarkonium-production 
rates at hadron colliders. The situation has slightly changed since then, with the first evaluations of the QCD 
corrections~\cite{Campbell:2007ws,Artoisenet:2007xi,Gong:2008sn,Gong:2008hk,Artoisenet:2008fc}
to the yields of  $J/\psi$ and $\Upsilon$ (commonly denoted $\Q$ hereafter) produced in high-energy 
hadron  collisions via Colour-Singlet (CS) transitions~\cite{CSM_hadron}. 
It is now indeed widely accepted~\cite{Lansberg:2008gk,ConesadelValle:2011fw,Brambilla:2010cs} that $\alpha^4_s$ and $\alpha^5_s$ corrections 
to the CSM are significantly larger than $\alpha^3_s$ contributions at mid and large $P_T$ and that they should be taken into 
account in any analysis of their $P_T$ spectrum. 
Nowadays, it not clear anymore that CO channels dominate and they are the only source of quarkonia. As a result, there is no consensus 
on which mechanisms are effectively at work in quarkonium hadroproduction at
high energies, that is at RHIC, at the Tevatron and, recently, at the LHC.

Polarisation predictions for the CS channel are also strongly affected by QCD corrections 
as demonstrated in~\cite{Gong:2008sn,Artoisenet:2008fc,Li:2008ym,Lansberg:2009db}.
At NLO, $\Q$  produced inclusively or 
in association with a photon are expected to be 
longitudinally polarised when $P_T$ gets larger, whereas they were thought to be transversely polarised
as predicted at LO in the CSM~\cite{Leibovich:1996pa,Kim:1994bm}. Such a drastic change is understood
 by the dominance of new production topologies. This
also explains the significant enhancement in the production rates as observed for increasing $P_T$.

The situation is rather different at low $P_T$, where  the CS predictions for $\Q$  at
LO~\cite{CSM_hadron}  and NLO~\cite{Campbell:2007ws,Artoisenet:2007xi,Gong:2008sn} 
 accuracy are of the same magnitude at RHIC energies; this shows a good convergence of the perturbative series. 
They are also in agreement~\cite{Brodsky:2009cf,Lansberg:2010cn,Lansberg:2012ta} with the existing data
from RHIC~\cite{Adare:2006kf} energy all the way up to that of the 
LHC~\cite{Abelev:2010am,Acosta:2001gv,Abazov:2005yc,Khachatryan:2010zg,Aad:2011xv,Aaij:2012ve,Aaij:2011jh,Aamodt:2011gj}.
CO channels are most likely not needed to account for low $P_T$ data --and thus for the $P_T$ integrated
yields. This is at odds with earlier works, \eg~\cite{Cooper:2004qe}, which wrongly assumed that 
$\chi_c$ feed-down could be the dominant CSM contribution.
This is supported further by the results of recent works~\cite{ee} focusing on production at $e^+ e^-$ 
colliders which have posed stringent constraints on the size of $C=+1$ CO contributions which 
can be involved in hadroproduction at low $P_T$. Finally, this is
reminiscent of the broad fixed-target
measurement survey of total cross sections~\cite{Maltoni:2006yp} which challenged the universality of
the CO MEs.

\begin{figure}[htb!]
\centering
\subfloat[Born]{\includegraphics[width=0.2\textwidth]{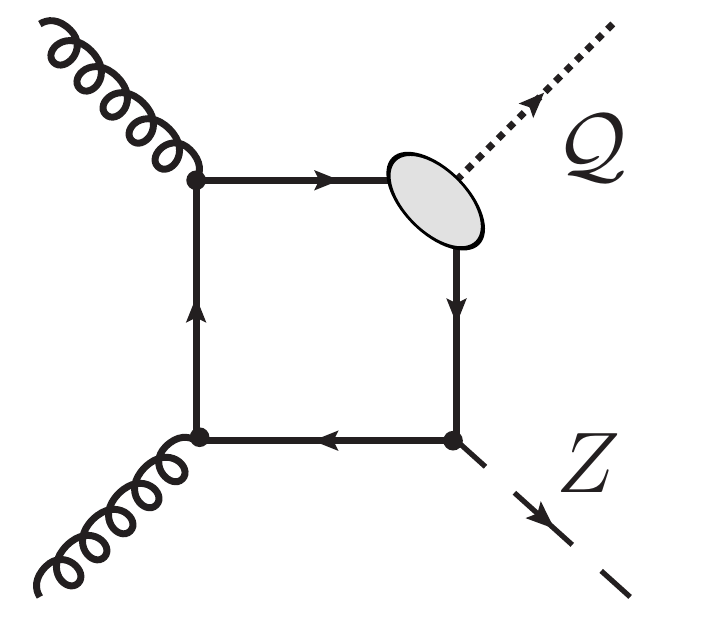}}\hspace*{-.2cm}
\subfloat[NLO loop ]{\includegraphics[width=0.2\textwidth]{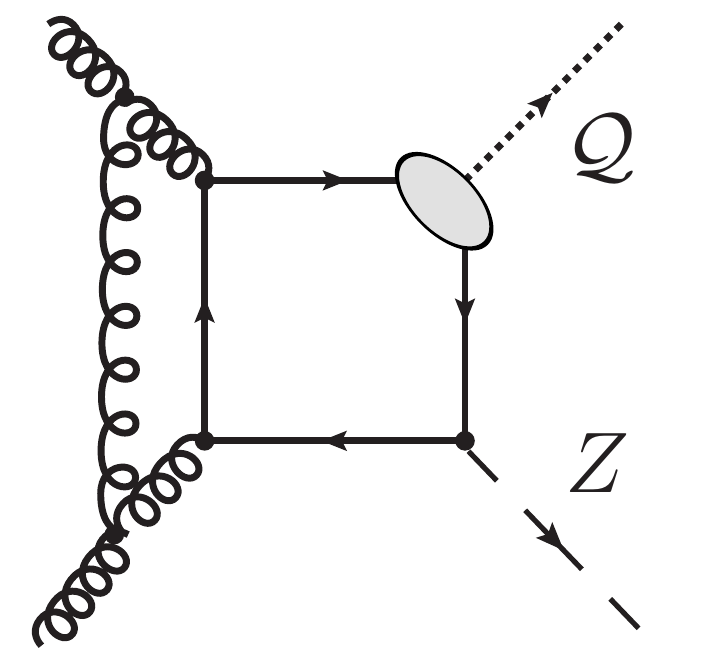}}\hspace*{-.2cm}
\subfloat[NLO real emission from the heavy quark]{\includegraphics[width=0.2\textwidth]{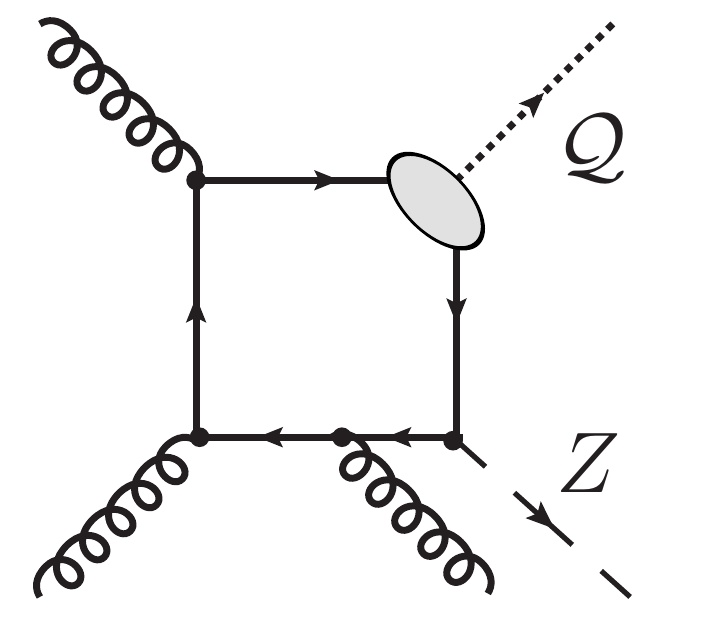}}\hspace*{-.2cm}
\subfloat[$t-$CGE]{\includegraphics[width=0.2\textwidth]{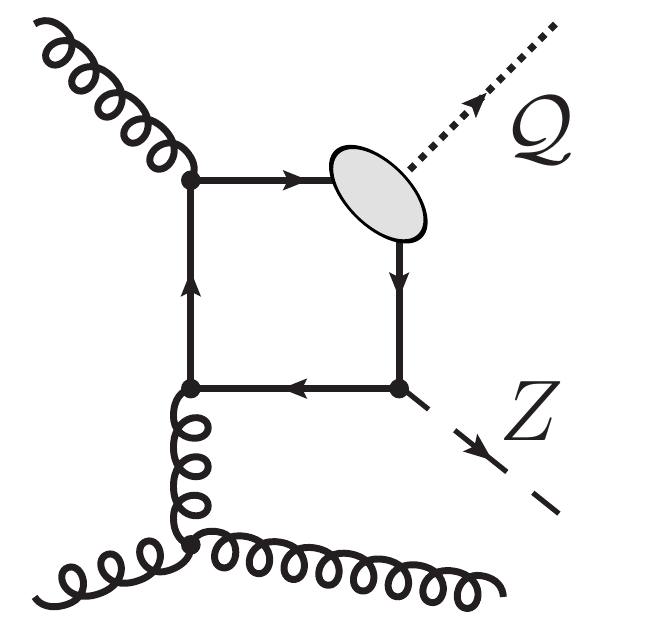}}\hspace*{-.2cm}
\subfloat[$t-$CGE]{\includegraphics[width=0.2\textwidth]{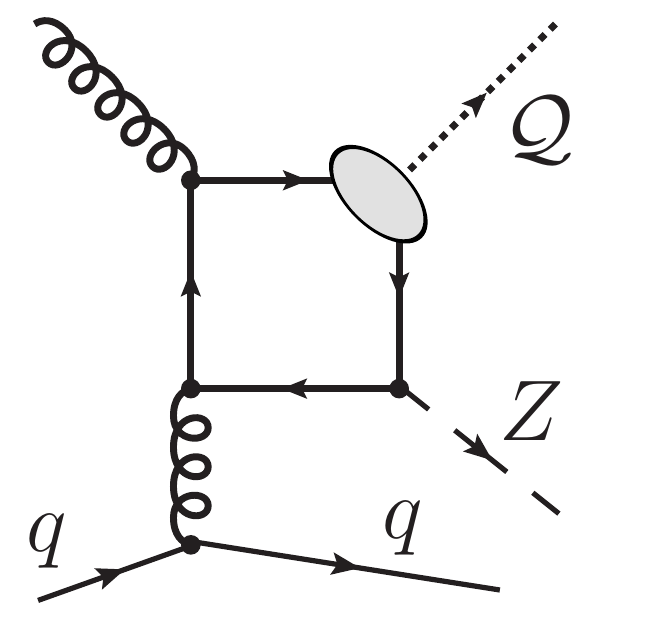}}\hspace*{-.2cm}
\caption{Representative diagrams contributing to $J/\psi$ and $\Upsilon$ {(denoted $\Q$)} hadroproduction with a $Z$ boson in the CSM
 by gluon fusion at orders $\alpha \alpha_s^2$ (a), $\alpha \alpha_s^3$ (b,c,d) and initiated
by a light-quark gluon fusion at order $\alpha \alpha_s^3$ (e).
The quark and antiquark attached to the ellipsis are taken as on-shell
and their relative velocity $v$ is set to zero.}
\label{diagrams}
\end{figure}

In this paper, we focus on the production of $J/\psi$ (and $\Upsilon$) in association with a $Z$ boson. Whereas this process may
give us complementary information on quarkonium production if it happens to be experimentally accessible at the LHC, it
also offers an interesting theoretical playground for the understanding of the QCD corrections in 
quarkonium-production processes. Our motivation was twofold: first, to see if the polarisation pattern of the 
$J/\psi$ is altered by the QCD corrections at large $P_T$; second, to see how large the effect of 
new topologies opening at NLO is, by comparing a full NLO computation to a simplified one --NLO$^\star$-- 
with a infrared (IR) cut-off and neglecting loops. Our attention has also been drawn to this process by a previous
analysis of the yield at NLO~\cite{Mao:2011kf}  which showed an intriguing result where NLO corrections were large 
at low $P_T$ and getting smaller at large(r) $P_T$. Such a result could only be explained by
a negligible contribution from new kinematically enhanced topologies and a large (positive) contribution
from loop corrections at low $P_T$. As we shall demonstrate, the conclusion drawn in~\cite{Mao:2011kf}
are misguided by an unconventional choice of the factorisation and renormalisation scales ($\mu_F$ and $\mu_R$), --way
below $m_Z$-- and a $P_T$ range not large enough --compared to $m_Z$-- to be able to observe the dominance of 
$t-$CGE topologies. As a matter of fact, if one chooses a value for the scales commensurate with $m_Z$, rather 
than the transverse  mass of the $J/\psi$ as done in~\cite{Mao:2011kf},  the NLO corrections are found to be small 
at small $P_T$. On the other hand, for $P_T\,\gsim\, m_Z/2$, the NLO corrections are enhanced by a kinematical factor $P_T^2$.

The paper is organised as follows. In sections \ref{sec:LO} and  \ref{sec:NLO}, we describe the evaluation 
of the cross section at LO and NLO accuracy in the CSM. We also explain how the partial 
NLO$^\star$ yield is evaluated. In section \ref{sec:results}, we present our 
results which we first compare to those from~\cite{Mao:2011kf} with the same scale choice, at the same energy 
and in the same kinematical region. Then, we show our predictions
in an extended $P_T$ range for $\mu_{F,R}$ commensurate with $m_Z$ and we discuss the ratio NLO over LO. We also study
the sensitivity of our prediction on the aforementioned scales.  Afterward, we compare 
the NLO$^\star$ yield with the full NLO and we comment on the dependence on the IR cut-off at large $P_T$ and on 
the impact of the $t-$CGE topologies. In section~\ref{sec:pol},  we analyse the yield polarisation at LO, NLO and NLO$^\star$.
In section~\ref{sec:upsilon}, we give and discuss our predictions for $\Upsilon$.  Section \ref{sec:conclusion} gathers our conclusions.

\section{Cross section at LO accuracy }\label{sec:LO}

In the CSM~\cite{CSM_hadron}, the matrix element to create
a $^3S_1$  quarkonium ${\Q}$ with a momentum $P_\Q$ and a polarisation $\lambda$
 accompanied by other partons, noted $j$, and a $Z$ boson of momentum $P_Z$
is the product of the amplitude to create the corresponding heavy-quark pair, ${\cal M}(ab \to Q \bar Q)$, a spin
 projector $N(\lambda| s_1,s_2)$ and $R(0)$, the radial wave function at the origin in the configuration
space, obtained from the leptonic width, namely 
\eqs{ \label{CSMderiv3}
{\cal M}&(ab \to {\Q}^\lambda(P_\Q)+Z(p_Z)+j)=\!\sum_{s_1,s_2,i,i'}\!\!\frac{N(\lambda| s_1,s_2)}{ \sqrt{m_Q}} \frac{\delta^{ii'}}{\sqrt{N_c}} 
\frac{R(0)}{\sqrt{4 \pi}}\\ &\times 
{\cal M}(ab \to Q^{s_1}_i \bar Q^{s_2}_{i'}(\mathbf{p}=\mathbf{0}) +Z(p_Z) + j),
}
where $P_\Q=p_Q+p_{\bar Q}$, $p=(p_Q-p_{\bar Q})/2$, $s_1$ and $s_2$ are the heavy-quark spins, and $\delta^{ii'}/\sqrt{N_c}$ 
is the projector onto a CS state. $N(\lambda| s_1,s_2)$ can be written as 
$ \frac{\ep^\lambda_{\mu} }{2 \sqrt{2} m_Q } \bar{v} (\frac{\mathbf{P_\Q}}{2},s_2) \gamma^\mu u (\frac{\mathbf{P_\Q}}{2},s_1) \,\, $ in the non-relativistic limit 
with   $\ep^\lambda_{\mu}$ being the polarisation vector of the quarkonium. Summing over the quark spin yields to traces
which can be evaluated in a standard way.

At LO, there is only a single partonic process at work, namely $gg\to J/\psi Z$ --completely analogous to
$gg\to J/\psi \gamma$ for $J/\psi$-prompt photon associated production-- with 4 Feynman graphs to be evaluated. 
One of them  is drawn on \cf{diagrams} (a). The differential partonic cross section is readily obtained from the amplitude
squared\footnote{The momenta of the initial gluons, $k_{1,2}$, are, as usual in the parton model, related to those of the colliding hadrons ($p_{1,2}$) through
$k_{1,2}=x_{1,2} \, p_{1,2}$. One then defines the Mandelstam variables for the partonic system: $\hat s = s x_1 x_2$, 
$\hat t=(k_1-P_{J/\psi})^2$ and $\hat u=(k_2-P_{J/\psi})^2$.}, 
\be \frac{d \hat \sigma}{d \hat t} = \frac{1}{16 \pi \hat s} \left| {\cal M}\right|^2, \ee
from which one obtains the double differential cross section in $P_T$ ($P_T\equiv P_{J/\psi,T}$)  and the $J/\psi$ rapidity, $y$, for $pp \to J/\psi Z$ after convolution
with the gluon PDFs and a change of variable:
\be 
\frac{d\sigma}{dydP_T}=\int_{x_1^{\rm min}}^1 dx_1 \frac{2 \hat s P_T g(x_1,\mu_F) g(x_2(x_1),\mu_F)}
{\sqrt{s}(\sqrt{s} x_1-m_T e^{y})}
\frac{d\hat \sigma}{d\hat t},
\ee
where $x_1^{\rm min}= \frac{m_T\sqrt{s}e^{y}-m_{J/\psi}^2+m_Z^2}{\sqrt{s}(\sqrt{s}-m_T e^{-y})}$, $m_T=\sqrt{m_{J/\psi}^2+P_T^2}$.

\section{Cross section at NLO accuracy} \label{sec:NLO}
The NLO contributions can be divided in two sets: one gathers the virtual corrections which arise from loop
diagrams, the other gathers the real (emission) corrections where one more particle appears in the final state.
In the next sections, we briefly describe how these are computed.

\subsection{Virtual corrections}\label{subsec:NLO-virt}

The computation of the virtual corrections involves three types of singularities: the ultraviolet (UV), the infrared (IR) and the Coulomb 
ones. UV divergences arising from self-energy and triangle
diagrams are cancelled after renormalisation. A similar renormalisation scheme as in Ref.~\cite{Gong:2009ng} is used, 
except for the fact that, in the present study, the bottom quark is also included in the renormalisation of the gluon field. 
The renormalisation constants $Z_m$, $Z_2$ and $Z_3$ which are associated to the charm quark mass $m_c$, the 
charm-field $\psi_c$ and the gluon field $A^a_{\mu}$ are defined in the on-mass-shell 
(OS) scheme while $Z_g$, for the QCD gauge coupling constant
$\alpha_s$ is defined in the modified minimal-subtraction ($\overline{\mathrm{MS}}$) scheme: \bea
\delta Z_m^{OS}&=&-3C_F\dfrac{\alpha_s}{4\pi}\left[\dfrac{1}{\e_{\rm UV}} -\gamma_E +\ln\dfrac{4\pi \mu_R^2}{m_c^2} +\frac{4}{3}\right] ,\NO \\
\delta Z_2^{OS}&=&-C_F\dfrac{\alpha_s}{4\pi}\left[\dfrac{1}{\e_{\rm UV}} +\dfrac{2}{\e_{\rm IR}} -3\gamma_E +3\ln\dfrac{4\pi
\mu_R^2}{m_c^2} +4 \right] ,\NO \\
\delta Z_3^{OS}&=&\dfrac{\alpha_s}{4\pi}\biggl[(\beta'_0-2C_A)\left(\dfrac{1}{\e_{\rm UV}} -\dfrac{1}{\e_{\rm IR}}\right)
\NO\\&&
-\dfrac{4}{3}T_F\left(\dfrac{1}{\e_{\rm UV}} -\gamma_E +\ln\dfrac{4\pi \mu_R^2}{m_c^2}\right) -\dfrac{4}{3}T_F\left(\dfrac{1}{\e_{\rm UV}} -\gamma_E +\ln\dfrac{4\pi \mu_R^2}{m_c^2}\right) \biggr] , \\\NO
\delta
Z_g^{\overline{\mathrm{MS}}}&=&-\dfrac{\beta_0}{2}\dfrac{\alpha_s}{4\pi}\left[\dfrac{1}{\e_{\rm UV}}
-\gamma_E +\ln(4\pi)\right] \NO, \eea 
where  $\g_E$ is Euler's constant, $\b_0=\frac{11}{3}C_A-\frac{4}{3}T_Fn_f$ 
is the one-loop coefficient of the QCD beta function and $n_f$ is the number of active quark flavours. 
We take the three light quarks $u, d, s$ as massless and consider the quarks $c$ and $b$ as heavy; therefore $n_f$=5. 
In $SU(3)_c$,  we have the following colour factor: $T_F=\frac{1}{2}, C_F=\frac{4}{3}, C_A=3$. Finally,
$\b'_0\equiv\b_0+\frac{8}{3}T_F=\frac{11}{3}C_A-\frac{4}{3}T_Fn_{lf}$ where $n_{lf}\equiv n_f-2=3$ is the 
number of light quark flavours.

After having fixed our renormalisation scheme, there are 111 virtual-correction diagrams, including counter-term diagrams.
Diagrams that have a virtual gluon line connecting the charm quark pair forming the $J/\psi$ lead to Coulomb 
singularity $\sim \pi^2/|p|$, which can be isolated and mapped into the $c\bar{c}$ wave function.

The loop integration has been carried out thanks to the newly upgraded Feynman Diagram Calculation (FDC) 
package~\cite{Wang:2004du}, with the implementation of the reduction method for loop integrals 
proposed in Ref.~\cite{Duplancic:2003tv}.

\subsection{Real corrections}\label{subsec:NLO-real}
The real corrections arise from three parton level subprocesses:
\bea
g+g&\rightarrow & J/\psi + Z + g, \label{eq:ggpsiZg} \\
g+q(\bar{q})&\rightarrow &J/\psi + Z + q(\bar{q}),\label{eq:gqpsiZq} \\
q+\bar{q}&\rightarrow & J/\psi + Z + g,\label{eq:qqpsiZg}
\eea
where $q$ denotes light quarks with different flavours ($u,d,s$). 
We have not considered the contributions from the processes $c\bar{c}\rightarrow J/\psi + Z +g$ and 
$g+c(\bar{c})\rightarrow J/\psi + Z + c\bar(c)$. Both are IR finite and can be safely separated out from the 
other ones. The charm-gluon fusion contribution may be non-negligible 
in the presence of intrinsic charm. It will be considered in a separate work.

The contribution from the quark-antiquark fusion (\ce{eq:qqpsiZg}) is also IR finite and small. 
The phase-space integration of the other two subprocesses will generate IR
singularities, which are either soft or collinear and which can be
conveniently isolated by slicing the phase space into different
regions. We use the two-cutoff phase-space-slicing method
\cite{Harris:2001sx}, which introduces two small cutoffs to
decompose the phase space into three parts. The real cross
section can then be written as \be
\s^{\rm Real}=\s^{\rm Soft}+\s^{\rm Hard\ Collinear}+\s^{\rm Hard\ Noncollinear} .\ee
The hard noncollinear part $\s^{\rm Hard\ Noncollinear}$ is IR finite and can be
numerically computed using standard Monte-Carlo integration techniques. 
Only the real subprocess of \ce{eq:ggpsiZg} contains soft singularities. 
Collinear singularities appear in both real subprocesses of \ce{eq:ggpsiZg} and \ce{eq:gqpsiZq}, 
but only as initial-state collinear singularities.
As shown in Ref.~\cite{Harris:2001sx}, all these singularities can be factored out analytically 
in the corresponding regions. When combined with the IR singularities appearing in the virtual corrections 
(see section~\ref{subsec:NLO-virt}), the soft singularities of the real part cancel. Yet, 
some collinear singularities remain. These are fully absorbed 
into the redefinition of the parton distribution function (PDF): this is usually referred to as the
 mass factorisation~\cite{Altarelli:1979ub}. All the singularities are thus eventually analytically cancelled.

\subsection{NLO$^\star$ cross section}\label{subsec:NLOstar}

In order to evaluate the NLO$^\star$ contributions, we use the framework described in~\cite{Artoisenet:2007qm} based on the
tree-level matrix-element generator {\small MADONIA}~\cite{Madonia} slightly tuned to implement an IR cut-off
on all light parton-pair invariant mass. The LO cross section has also been checked with {\small MADONIA}.

The procedure used here to evaluate the leading-$P_T$ NLO contributions is exactly the same as 
in~\cite{Artoisenet:2008fc} but for the process $pp\to J/\psi+Z+\hbox{jet}$. Namely, the 
real-emission contributions at $\alpha \alpha_s^3$ are evaluated using {\small MADONIA} by 
imposing a lower bound on the invariant mass of 
any light parton pair ($s^{\rm min}_{ij}$). The underlying idea in the inclusive\footnote{``Inclusive'' is used
here in opposition to ``in association with another detected particle'' which is indeed a more exclusive process.}   case 
was that for the new channels opening up at NLO which have a leading-$P_T$ behaviour (for instance the $t$-CGE), 
 the cut-off dependence should decrease for increasing $P_T$ since no collinear or soft divergences can appear there.
For other NLO channels, whose Born contribution is at LO, the cut would 
produce logarithms of $s_{ij}/s_{ij}^{\rm min}$, which are not necessarily negligible. Nevertheless, 
they can be factorised over their corresponding Born contribution, which scales as $P_T^{-8}$, and are thence 
suppressed by at least two powers of $P_T$ with respect to the leading-$P_T$ contributions ($P_T^{-6}$) at this order. 
The sensitivity on $s_{ij}^{\rm min}$ should vanish at large $P_T$. This argument 
has been checked in the inclusive case for $\Upsilon$~\cite{Artoisenet:2008fc} and $\psi$~\cite{Lansberg:2008gk}
  as well as in association with a photon~\cite{Lansberg:2009db}.
Because of the presence of the $Z$ boson mass, it is not a priori obvious that $t-$CGE topologies  dominate over
the LO ones. It is thus not clear at all how such procedure to evaluate
the NLO$^\star$ yield can provide a reliable evaluation of the full NLO of $J/\psi +Z$. In fact, at mid $P_T$, significantly below
the $Z$ boson mass, the difference of the $P_T$ dependence of the NLO and LO cross 
sections is maybe not large enough for the dependence on 
$s_{ij}^{\rm min}$ to decrease fast. Having at hand a full NLO computation, we can carry out such a
comparison and better investigate the effect of QCD corrections in quarkonium production. This is done after our
complete results are presented.

\section{Results for $J/\psi+Z$: differential cross section in $P_T$} \label{sec:results}

\subsection{Comparison with~Mao \etal~\protect\cite{Mao:2011kf}}

In order to compare our results with those of~\cite{Mao:2011kf}, we take $\sqrt{s}=14$ TeV and $|y^{J/\psi}|< 3.0$. We also
set the factorisation and renormalisation scales at the same value, namely $\mu_F=\mu_R=m_T^{J/\psi}=\sqrt{m_{J/\psi}^2+P_T^2}$. 
We also take $\alpha=1/137$, $|R_{J/\psi}(0)|^2=0.91$~GeV$^3$, $m_c=1.5$~GeV, $m_z=91.1876$~GeV and $\sin^2(\theta_W)=0.23116$.
Our LO results (also cross-checked with MADONIA) do match with those of~\cite{Mao:2011kf} (compare both blue curves 
on \cf{fig:comp-with-mao}).

\begin{figure}[htb!]
  \centering 
\includegraphics[width=0.55\textwidth]{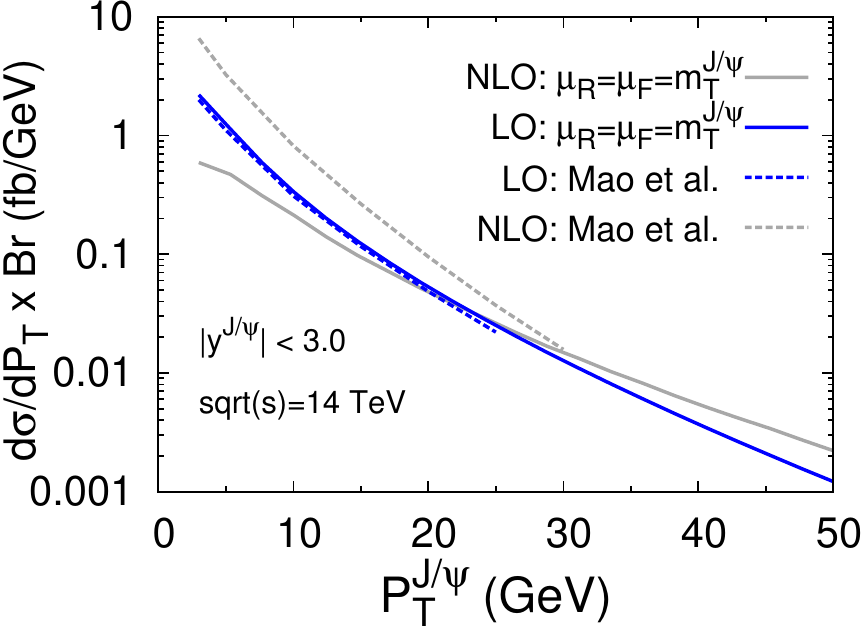}
\caption{Comparison between our results (solid lines) and that of Mao et al. ~\protect\cite{Mao:2011kf} (dashed lines) for the 
differential cross section for $J/\psi+Z$ vs. the $J/\psi$ $P_T$ at LO (blue) and NLO (gray) with $\mu_F=\mu_R=m_T^{J/\psi}$.} 
\label{fig:comp-with-mao}
\end{figure}

However, as depicted in \cf{fig:comp-with-mao}, we are not able to reproduce
the NLO results presented in the later reference. At low $P_T$, we have found a $K$ factor smaller than one 
(\ie~the yield at NLO is smaller than at LO) while they obtained a value larger than one.
The way $\alpha_s$ is precisely evaluated in both computations may differ but this can hardly explain a sign change
in the $\alpha \alpha_s^3$ contributions. There is also a difference in the renormalisation scheme: 
we have included both charm and bottom quarks in the renormalisation of the gluon field contrary to what has been done 
in the previous analysis. Yet, we do not believe that this could explain the discrepancies between both results.

That being said, a scale close to $m_Z$, rather than the transverse mass of the $J/\psi$ taken in~\cite{Mao:2011kf}, seems 
more appropriate as done for instance for $Z+b-$jet~\cite{Campbell:2003dd}. 
This has an important effect on the scale sensitivity, less 
on the final numbers predicted for the yields, 
as we shall discuss in the next section.

\begin{figure}[htb!]
  \centering 
\subfloat[Scale dependence with $\mu_0=m_T^{J/\psi}$]{\includegraphics[width=0.5\textwidth]{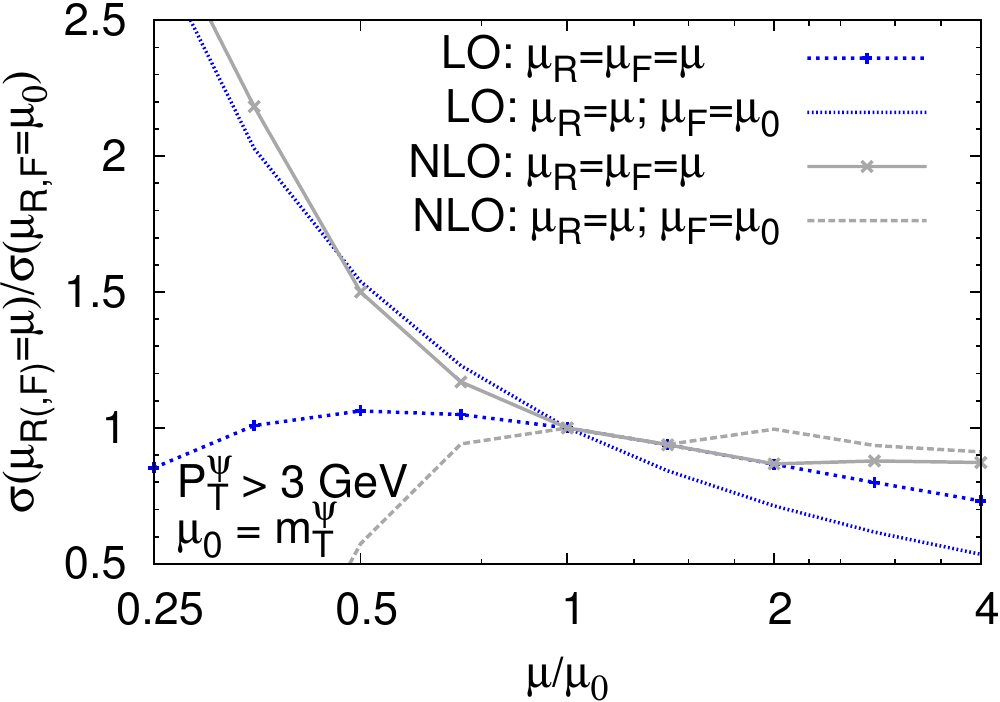}}
\subfloat[Scale dependence with $\mu_0=m_Z$]{\includegraphics[width=0.5\textwidth]{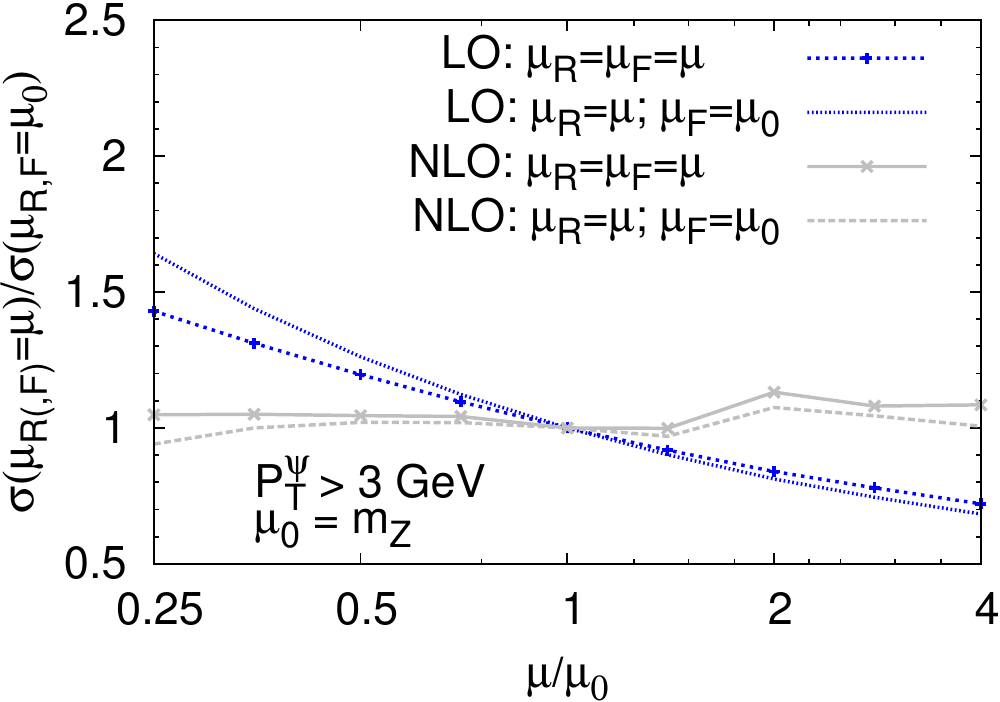}}
\caption{(a) Renormalisation and factorisation scale dependence of 
the LO and NLO yield for $P_T>3$ GeV with $\mu_0=m_T^{J/\psi}$. (b) Same plot as (a) for $\mu_0=m_Z$.}
\label{fig:comp-mT-mZ}
\end{figure}

In \cf{fig:comp-mT-mZ}, we show the scale sensitivity at low $P_T$ around two different choices
of the ``default'' scale value, $\mu_0$, (a) the transverse mass of the $J/\psi$ and (b)  the $Z$ boson mass. We emphasise
that we believe the latter choice to be more appropriate owing to the presence of the $Z$ boson in the hard process. 
One sees that around $m_Z$ (b), the cross section at NLO is more stable, 
except for the bump at $2m_Z$ which can be corrected by properly setting the value of $\Lambda^{[6]}$ in the running 
of coupling constant (currently 0.151 MeV with $m_t=180$~GeV), which matters for $\mu_R > m_t$. The NLO results
are clearly unstable at low scales and they may then artificially be enhanced. In the following sections, we investigate
further the dependence of the scale sensitivity for different domains  of the $J/\psi$ transverse momenta.

\subsection{Results for the differential cross section in $P_T$ at $\sqrt{s}=8$ TeV and 14 TeV}

In the following, we show our results  for $|y^{J/\psi}|< 2.4$ --the usual $J/\psi$ acceptance for the CMS and ATLAS detectors--
at 8 TeV and\footnote{The cross section at 13 TeV is 12 \% smaller than at 14 TeV.} 14 TeV and for the 
renormalisation and factorisation scales set at $m_Z$. We have 
kept\footnote{Note that we could have evaluated the cross section for lower $P_T$ where the cross section is well behaved. 
However, we do not expect --at least in the central region-- any experimental measurement 
to be carried out in this region owing to the momentum cut on the muons because of the strong magnetic fields in the ATLAS and CMS detectors.} 
the cut $P_T^{J/\psi} > 3$ GeV.

The parameters entering the cross-section evaluation have been taken as follows: $|R_{J/\psi}(0)|^2=0.91$ GeV$^3$, 
Br$(J/\psi \to \ell^+\ell^-)=0.0594$, $m_c=1.5$ GeV with $m_{J/\psi}=2m_c$, $m_b=4.75$ GeV,  
$\alpha=1/128$. Our result at $\sqrt{s}=14$ TeV are depicted
on \cf{fig:dsigpdt-NLO-14TeV}. The dotted blue line is our LO result and the solid gray line is our prediction at NLO. It is 
obvious, contrary to what was obtained in~\cite{Mao:2011kf}, that the yield at NLO is getting larger than at LO for increasing
$P_T$. This is similar to what happens in the inclusive case. This is also indicative that new leading $P_T$ topologies, 
in particular $t$-CGE, start dominating rather early in $P_T$ despite of the presence of a $Z$ boson in the process. 
At $P_T =150$~GeV, the NLO yield is already ten times that of LO. 

\begin{figure}[htb!]
\centering
{\includegraphics[width=0.65\textwidth]{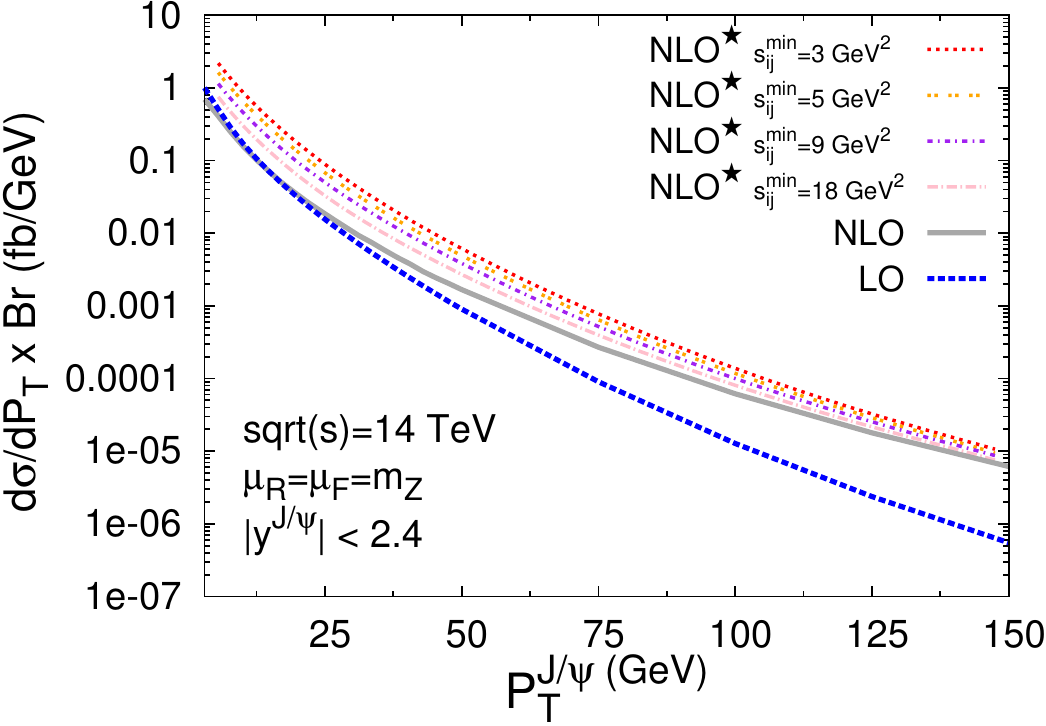}}
\caption{Differential cross section for $J/\psi+Z$ vs. $P_T$ at $\sqrt{s}=14$ TeV at LO (blue dashed) and NLO (gray solid)
with  $\mu_F=\mu_R=m_Z$ along with the NLO$^\star$ for different values of $s_{ij}^{\rm min}$ (red dotted, yellow double dotted, 
purple dash dotted and pink long-dash dotted). 
}
\label{fig:dsigpdt-NLO-14TeV}
\end{figure}

The dominance of $t$-CGE topologies can be quantified by a comparison with the results from the 
NLO$^\star$ evaluation. As aforementioned, because of the $Z$ boson mass, it was not a priori clear 
that  the NLO$^\star$ evaluation could make any sense here.  Indeed, as long as the contribution 
from the sub-leading $P_T$ topologies are significant, the NLO$^\star$ would strongly depend on 
the arbitrary IR cutoff\footnote{Not to be confused with the cutoff used in the full NLO 
computation, on which the final results does not depend.}  which is used to mimic the effect of the loop 
contributions which regulate the soft gluon emission divergences. We are in a position to check 
from which $P_T$ the NLO$^\star$ starts to reproduce the full NLO and becomes to be less sensitive 
on the IR cut. 

The various dotted lines on \cf{fig:dsigpdt-NLO-14TeV} show the NLO$^\star$ evaluation for different 
cut-off values. Two observations can be made: 1) they converge to the NLO steadily for increasing 
$P_T$, 2) for $P_T> m_Z$, the NLO$^\star$ evaluations are within a factor of 2 compatible with the complete NLO
yield. This confirms that loop corrections are sub-leading in $P_T$ and can be safely neglected 
for $P_T$ larger than all the masses relevant for the process under consideration and that new 
topologies appearing at NLO, the $t$-CGE ones, dominate at large $P_T$. At low $P_T$, where the NLO and LO yield 
are similar, the NLO$^\star$ overestimate the NLO.

\begin{figure}[htb!]
\centering
{\includegraphics[width=0.6\textwidth]{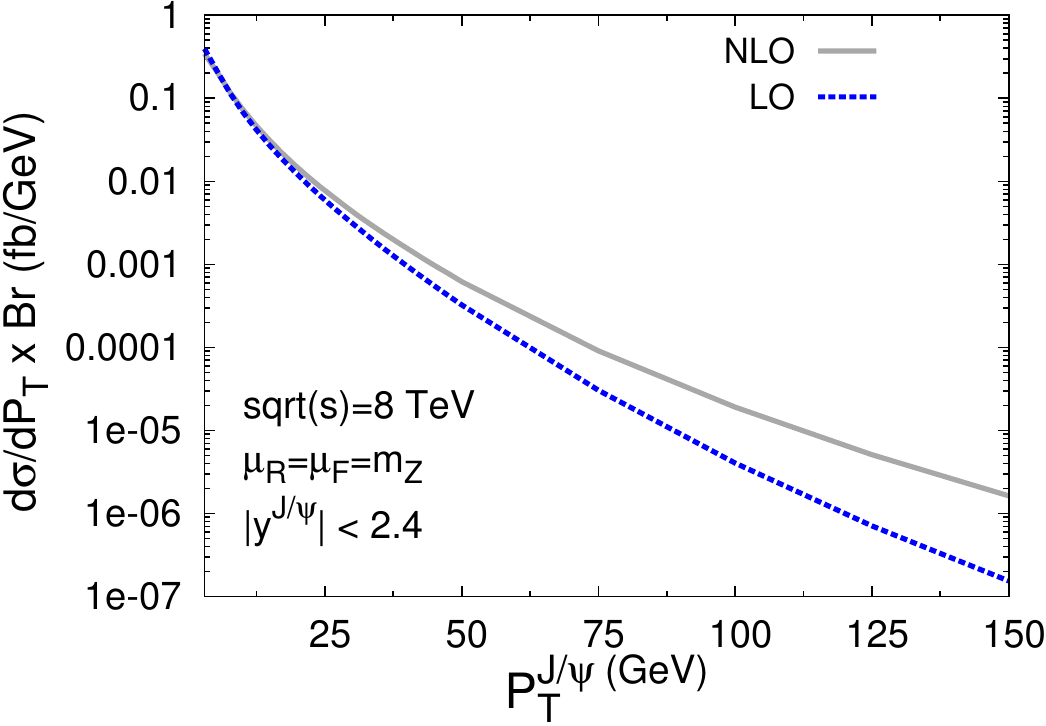}}
\caption{Differential cross section for $J/\psi+Z$ vs. $P_T$ at $\sqrt{s}=8$ TeV at LO (blue dashed) and NLO (gray solid)
with  $\mu_F=\mu_R=m_Z$. 
}
\label{fig:dsigpdt-NLO-8TeV}
\end{figure}

As regards the possibility to study such a process at the LHC, the $P_T$ differential cross sections 
times the branching ${\rm Br}(J/\psi \to \mu^+ \mu^-)$ at the smallest  $P_T$ accessible by ATLAS and CMS 
(3 to 5 GeV depending on the rapidity) is of the order of 1 fb/GeV at 
14 TeV (\cf{fig:dsigpdt-NLO-14TeV}) and three times less at 8 TeV (\cf{fig:dsigpdt-NLO-8TeV}). These
do not take into account the branching of the $Z$ in dimuons ($\sim 3\%$), but these are central values which
can be affected by a factor of 2-3 of theoretical uncertainties. In the most optimistic case, by integrating on the 
accessible $P_T$ range, by using both muon and electron decay channels for the $J/\psi$, by expecting
an indirect cross section of 40 $\%$ and by detecting the $Z$ boson with hadronic channels such that 
it could be detected 40 $\%$ of the time, it may be envisioned to detect something like four hundred events at 
14 TeV with 100 fb$^{-1}$ of data --possibly a factor 2-3 above given the theory uncertainty.
At 8 TeV with the 20 fb$^{-1}$ of data expected to be collected in 2012, we expect only about thirty events to be recorded.  
Clearly, there are more promising processes, such as $J/\psi+\gamma$~\cite{Li:2008ym,Lansberg:2009db} or $J/\psi+D$~\cite{Brodsky:2009cf,lhcb:2012dz}, to learn more on 
the production mechanisms of the $J/\psi$. Nevertheless, the study of $J/\psi+Z$ may suffer less from trigger
limitations and could thus still be  at reach at the LHC. In any case, it is an ideal theory playground 
to analyse the effects of QCD corrections on quarkonium production, which have been the key subject in the recent years
in the field.

\subsection{Scale sensitivity at different $P_T$}

From the observations made above, we expect the real emission contributions
 at $\alpha \alpha^3_s$ to dominate for $P_T\, \gtrsim\, m_Z/2$. 
This should therefore impact on the scale dependence of the yield. At low $P_T$ ($\ll m_Z$), 
we expect a reduced scale dependence since we really deal with a process at NLO accuracy. 
At large $P_T$, the leading process is  $pp \to J/\psi +Z+\hbox{parton}$. The loop contributions
are not expected to reduce the scale sensitivity since they are small. On the contrary, we expect a larger sensitivity
 on the renormalisation scale, $\mu_R$, since the leading process shows an additional power of $\alpha_s(\mu_R)$.
In practice, we  study the scale sensitivity by varying $\mu_F$ and $\mu_R$ together and then $\mu_R$ 
alone by a factor 2 about the ``default'' scale $m_Z$ with 3 cuts in $P_T$ --\ie~3, 50 and 150 GeV.

\begin{figure}[htb!]
\centering 
\subfloat[Scale dependence at LO]{\includegraphics[width=0.5\textwidth]{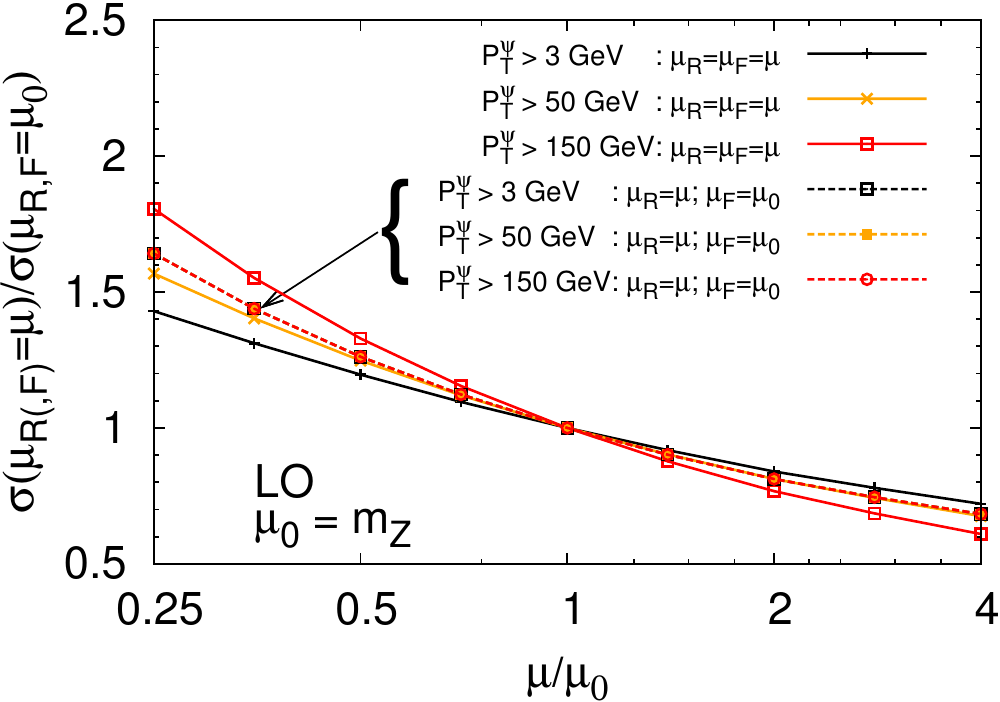}}
\subfloat[Scale dependence at NLO]{\includegraphics[width=0.5\textwidth]{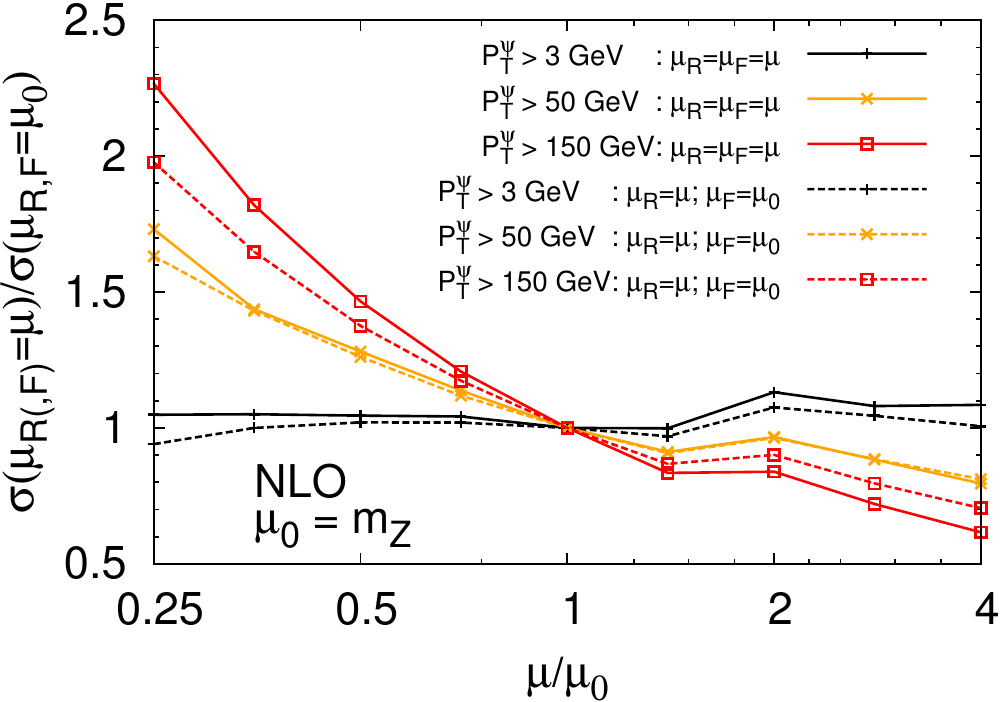}}
\caption{ Scale dependence of 
the yield at LO (a) and NLO (b) for $P_T>3$~GeV, $P_T>50$~GeV, $P_T>150$~GeV 
where both the renormalisation and factorisation scales are varied together ($\mu_F=\mu_R$, solid lines)
about  $\mu_0=m_Z$ and only the renormalisation scale is varied ($\mu_F$ fixed, dashed lines). Note that $\alpha$ 
has been kept fixed.}
\label{fig:scale-sensitivy}
\end{figure}

On~\cf{fig:scale-sensitivy},  we do observe, as anticipated for $P_T\,\gsim\, m_Z$ (red curves), a stronger scale sensitivity 
of the NLO yield (b) --at $\alpha \alpha_s^3$-- than of the LO yields (a)-- at $\alpha \alpha_s^2$. The NLO curve with $\mu_F$
fixed clearly shows that the sensitiviy essentially comes from $\mu_R$.
At mid $P_T$ (orange curves), the scale sensitivities are similar at LO and NLO, while at low $P_T$ (black curves), the NLO yield is
less scale dependent than the LO --in agreement with the common wisdom regarding the NLO computations. Note also that the 3 LO curves showing the
sole dependence of $\mu_R$ are identical since one can factor out a common $\alpha_s^2$ since our choices of $\mu_R$ do 
not depend on $P_T$.

\section{Polarisation: polar anisotropy in the helicity frame}\label{sec:pol}

The polar anisotropy of the dilepton decay of the $J/\psi$, $\lambda_\theta$ or $\alpha$, can be evaluated from
the polarised hadronic cross sections:
\be
\alpha(P_T)=\frac{\frac{d\s_T}{d P_T}-2 \frac{d\s_L}{d P_T}}
                 {\frac{d\s_T}{d P_T}+2 \frac{d\s_L}{d P_T}}.
\ee

To evaluate $\alpha(P_T)$, the polarisation of $J/\psi$ must of course be kept throughout 
the calculation. The partonic differential cross section for a polarised $J/\psi$ is expressed as: 
\be
\dfrac{d \hat{\s}_{\lambda}}{d \hat t}= a~\epsilon(\lambda) \cdot
\epsilon^*(\lambda) + \sum_{i,j=1,2} a_{ij} ~p_i \cdot
\epsilon(\lambda) ~p_j \cdot \epsilon^*(\lambda), \label{eq:def_pol_xsect}
\ee where $\lambda=T_1,T_2,L$.
$\epsilon(T_1),~\epsilon(T_2),~\epsilon(L)$ are respectively the two transverse
and the longitudinal polarisation vectors  of $J/\psi$; the polarisations of all the other particles are summed over in
n dimensions. One can find that $a$ and $a_{ij}$ are finite when the virtual  and real corrections are properly handled as
aforementioned. There is therefore no difference in the partonic differential cross section ${d \hat{\s}_{\lambda}}/{d \hat t}$
whether the polarisation of $J/\psi$ is summed over in 4 or $n$ dimensions. Thus, we can just treat the polarisation vectors of
$J/\psi$ in 4 dimensions.  There are usually several different choices of the polarisation 
frames, as discussed in Ref.~\cite{Beneke:1998re,Faccioli:2010kd,Faccioli:2012nv}. In our calculation, we have chosen to work
in the helicity frame. The polarisation can be obtained in a given frame by taking the corresponding polarisation vectors 
in~\ce{eq:def_pol_xsect}.

\begin{figure}[htb!]
\centering
\subfloat[$\sqrt{s}=14$ TeV]{\includegraphics[width=0.5\textwidth]{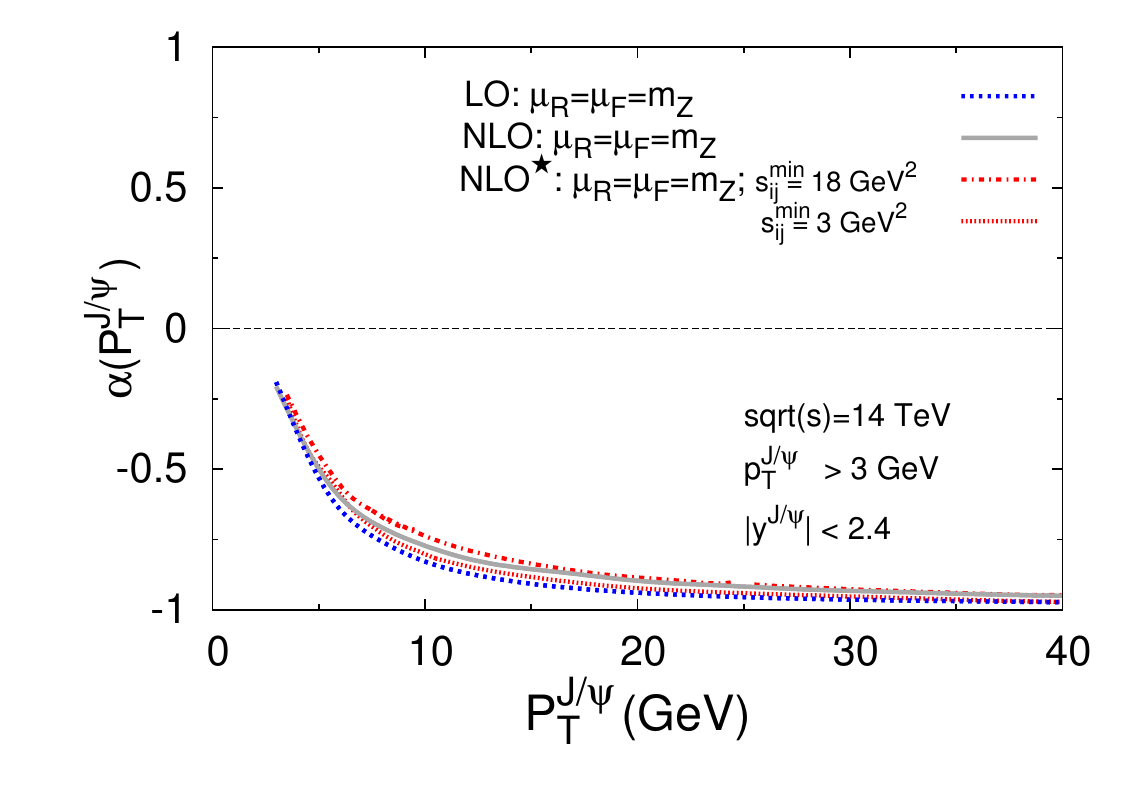}}
\subfloat[$\sqrt{s}=8$ TeV]{\includegraphics[width=0.5\textwidth]{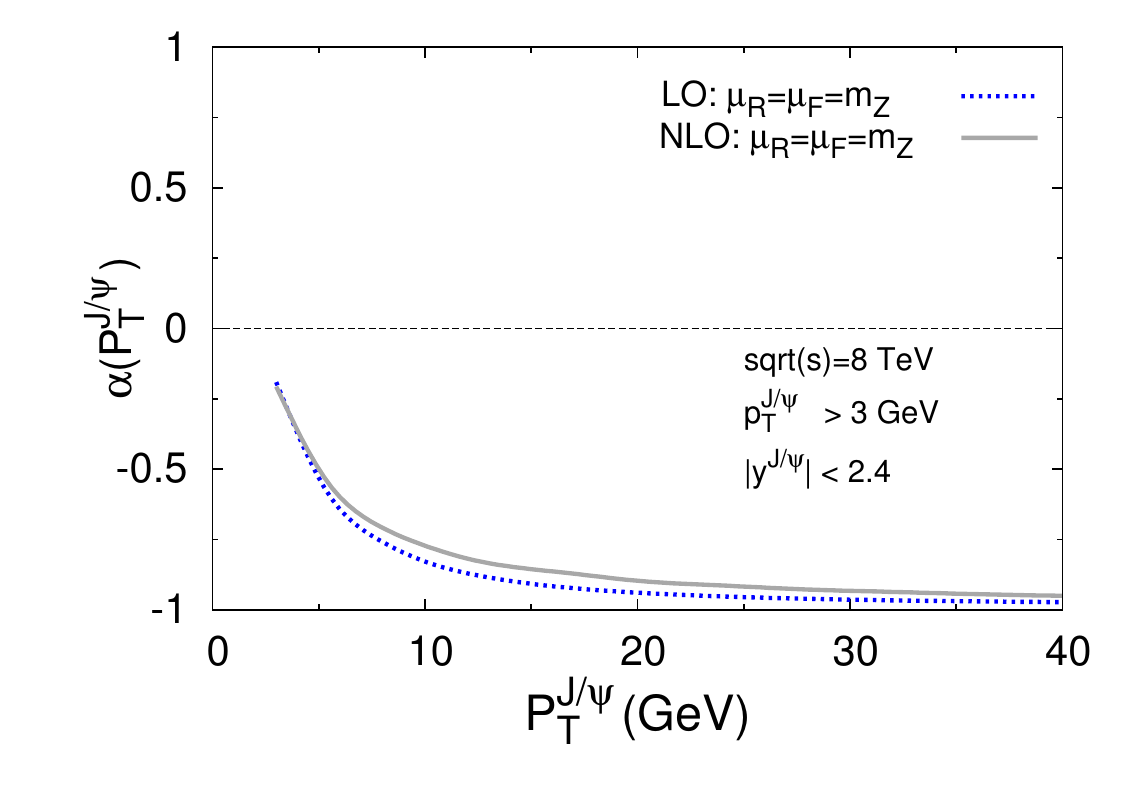}}
\caption{(a) $P_T$ dependence of the polarisation (or azimuthal anisotropy) in the helicity frame of
the  direct $J/\psi$ produced with a $Z$ boson at LO, NLO and NLO$^\star$ (for 2 values of the IR cut-off) at $\sqrt{s}=14$ TeV. (b) Same as (a) at LO and NLO at $\sqrt{s}=8$ TeV.}
\label{fig:pol-jpsi}
\end{figure}

Our results at 14 TeV in \cf{fig:pol-jpsi} (a) clearly show that the direct-$J/\psi$ yield in association with a $Z$ boson
is increasingly longitudinally polarised in the helicity frame for increasing $P_T^{J/\psi}$. The NLO and 
NLO$^\star$ results coincide and the latter is nearly insensitive to the IR cutoff. Interestingly, the NLO and the 
LO results are also very similar. This is the first time that such a robustness of the polarisation 
against QCD corrections is observed for the colour-singlet channels. For the $J/\psi$ produced inclusively 
or in association with a photon, 
the yield at LO and NLO are found to have a completely different polarisation. Our interpretation is that, 
when a $Z$ boson is emitted by one of the charm quarks forming the $J/\psi$, the latter is longitudinally 
polarised, irrespective of the off-shellness and of the transverse momentum of the gluons producing the 
charm-quark pair. This is not so when a photon or a gluon is emitted in the final
state. In the present case, we also note that 
the polarisation at 8 TeV (\cf{fig:pol-jpsi} (b)) is nearly exactly the same as at 14 TeV.

\section{Results for $\Upsilon+Z$}\label{sec:upsilon}

Along the same lines as for $J/\psi$, we have also evaluated the cross section and the polarisation
for direct-$\Upsilon$ production in association with a $Z$ boson. Experimentally, the CDF Collaboration
at Fermilab has set a 95 \% C.L. upper value for such a cross section at $\sqrt{s}=1.8$~TeV~\cite{Acosta:2003mu}, namely
\eqs{\sigma(p\bar p \to \Upsilon+Z +X) \times \hbox{Br}(\Upsilon\to \mu^+ \mu^-) < 2.5 \hbox{ pb}.} 

Further studies 
with the entire data set recorded by CDF is under process~\cite{private}. At $\sqrt{s}=1.8$ TeV, a quick evaluation
of the total cross section (without $y$ cut, nor $P_T$ cut) gives, for the CSM, a value close\footnote{We should 
however keep in mind that these are central values and that theoretical uncertainties can be 
of the order of a factor 2-3.} to 0.1 fb ($\sim 0.2$~fb by taking into account a similar feed-down 
fraction ($\sim 50 \%$) than that for the inclusive case). A similar evaluation for the
CO transitions is highly dependent on the choosen LDME values  and on the expected impact of the feed-down. Values span
from $\sim 0.06$ fb for the direct yield with CO LDMEs fit~\cite{Cho:1995ce} from the early prompt Tevatron data, up to
$\sim 3.75$ fb as evaluated in~\cite{Braaten:1998th}, passing by $\sim 0.4$ fb for the direct yield using CO LDMEs fit from
the latest Tevatron results taking into account some NLO QCD corrections~\cite{Gong:2010bk}.

\begin{figure}[htb!]
\centering\hspace*{-0.375cm}
\subfloat[$\sqrt{s}=14$ TeV]{\includegraphics[width=0.36\textwidth]{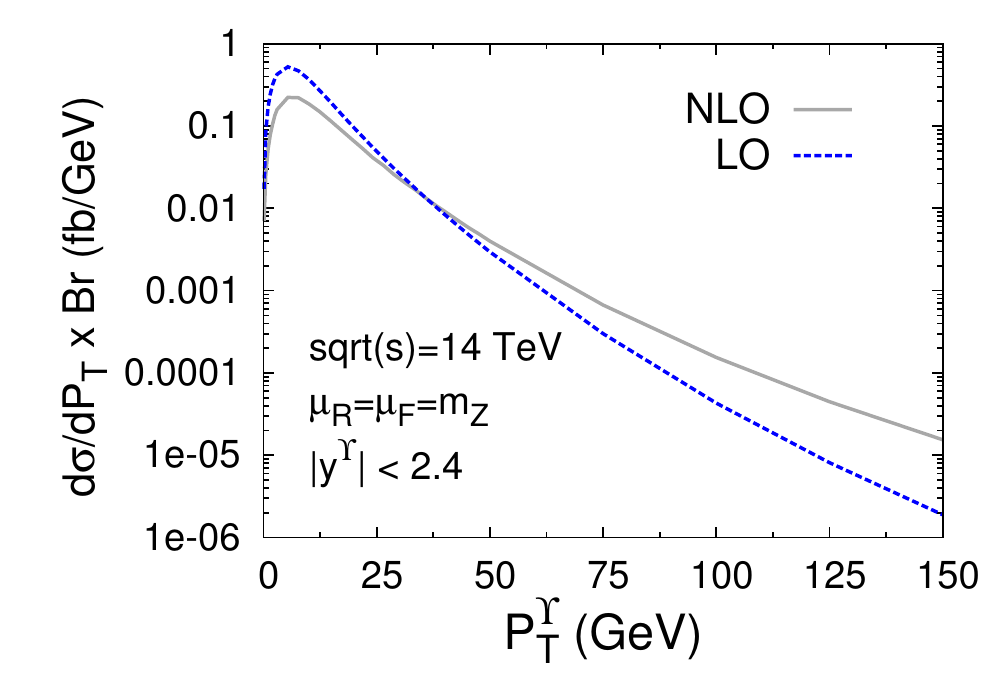}}\hspace*{-0.5cm}
\subfloat[$\sqrt{s}=8$ TeV]{\includegraphics[width=0.36\textwidth]{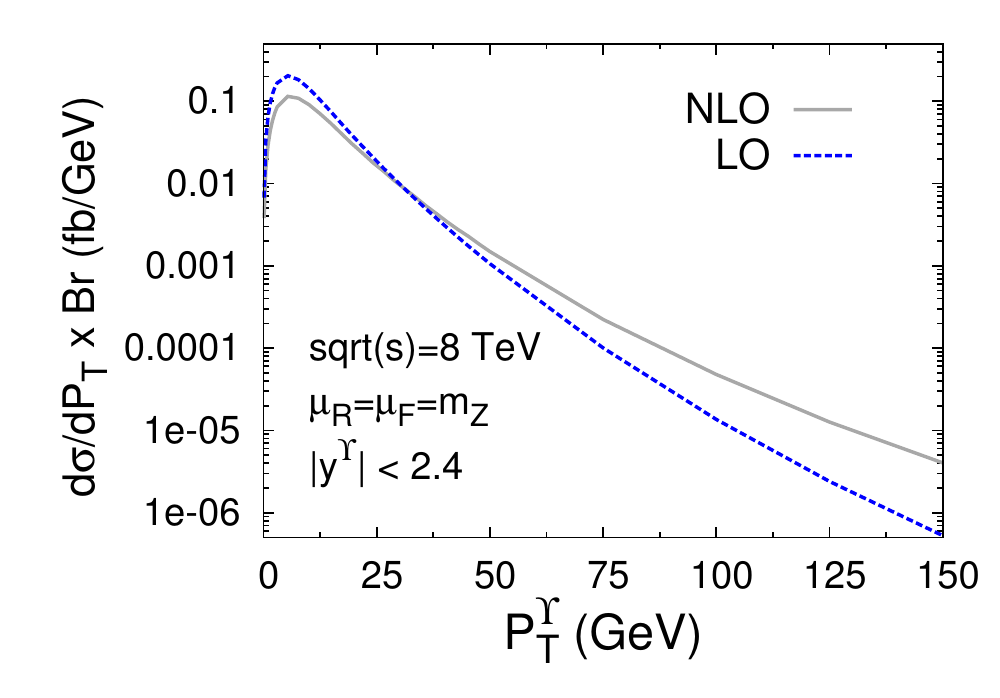}}\hspace*{-0.5cm}
\subfloat[$\sqrt{s}=1.96$ TeV]{\includegraphics[width=0.36\textwidth]{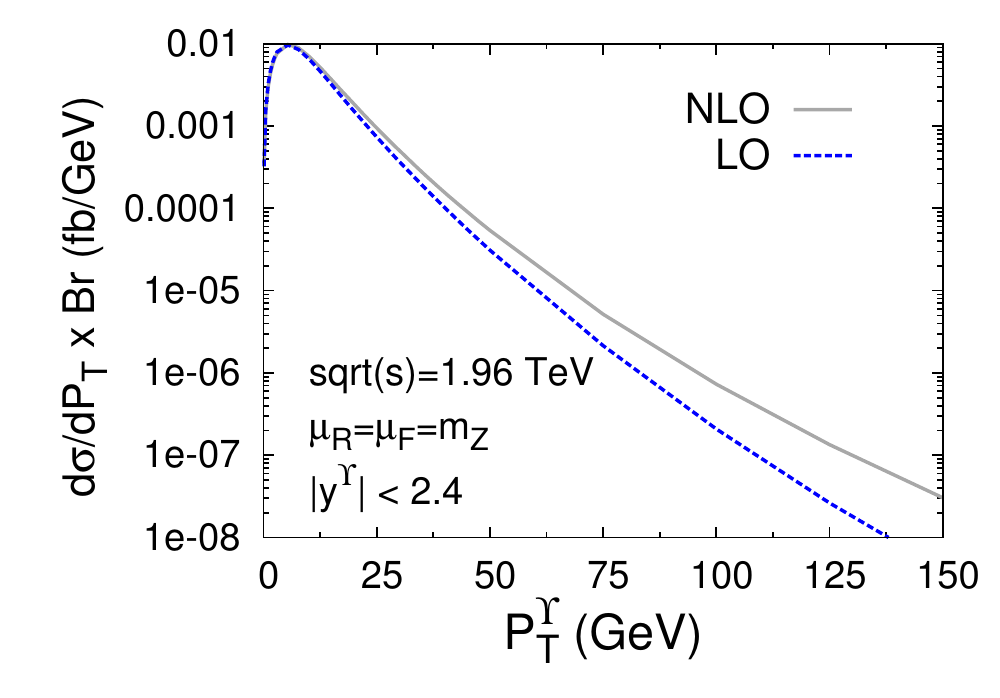}}
\caption{Differential cross section for direct $\Upsilon+Z$ vs. $P_T$ at LO (blue-dashed) and NLO (gray-solid)
with  $\mu_F=\mu_R=m_Z$ at 14 TeV (a), 8 TeV (b) and\protect\footnotemark\,  1.96 TeV (c).}
\label{dsdpt-Upsilon}
\end{figure}

\footnotetext{We have considered a wider rapidity range
than usual for the CDF quarkonium analyses since CMX muons can be used in such a correlation analysis owing to
the smaller background compared to inclusive measurements~\cite{private}.}

 This, is in any case, significantly below the CDF upper bound obtained with 83 pb$^{-1}$ of data. 
Given these small theoretical values, we fear that such process cannot be experimentally accessed at the Tevatron, 
unless contributions from colour-octet transitions, from  double-parton interactions or from 
feed-downs are unexpectedly large. At the LHC at 14 TeV, 
the expected yield in the CSM for the central rapidity region accessible by CMS and ATLAS 
is of the order 5~fb (still including the branching of the $\Upsilon$ in muons). The 
central values for the differential cross sections vs. $P_T$ at LO and NLO are shown in~\cf{dsdpt-Upsilon} (a-c). An enhancement 
by a factor 2 to 4 can certainly be expected if the feed-downs from excited bottomonium states and the usual 
theoretical uncertainties are taken into account. 

By comparing~\cf{dsdpt-Upsilon} (a-c), one also notices an interesting phenomenon: the NLO and LO yields 
start to depart from each other at low $P_T$ for increasing $\sqrt{s}$. This can probably be attributed to 
an increasing --negative-- size of the loop corrections in this region at small $x$. This is in fact reminiscent
to what has been observed in the inclusive case~\cite{Campbell:2007ws,Brodsky:2009cf,Gong:2008sn,Lansberg:2010cn}. 
In the latter case, the situation is worse since the NLO cross section can become negative for large  $\sqrt{s}$ 
and small $P_T$. It hints at significant NNLO corrections at small $P_T$ in the small-$x$ region, as anticipated in~\cite{Khoze:2004eu}.

For the sake of the
comparison with the $J/\psi$ case, we have also computed the polarisation at LO and NLO. 
As it can be seen on \cf{fig:pol-upsilon}, the yield polarisation at LO and NLO are very alike, 
though slightly different from for the $J/\psi$ case --most probably
due to the change in the quarkonium mass compared to the $Z$ mass.

\begin{figure}[htb!]
\centering
\includegraphics[width=0.6\textwidth]{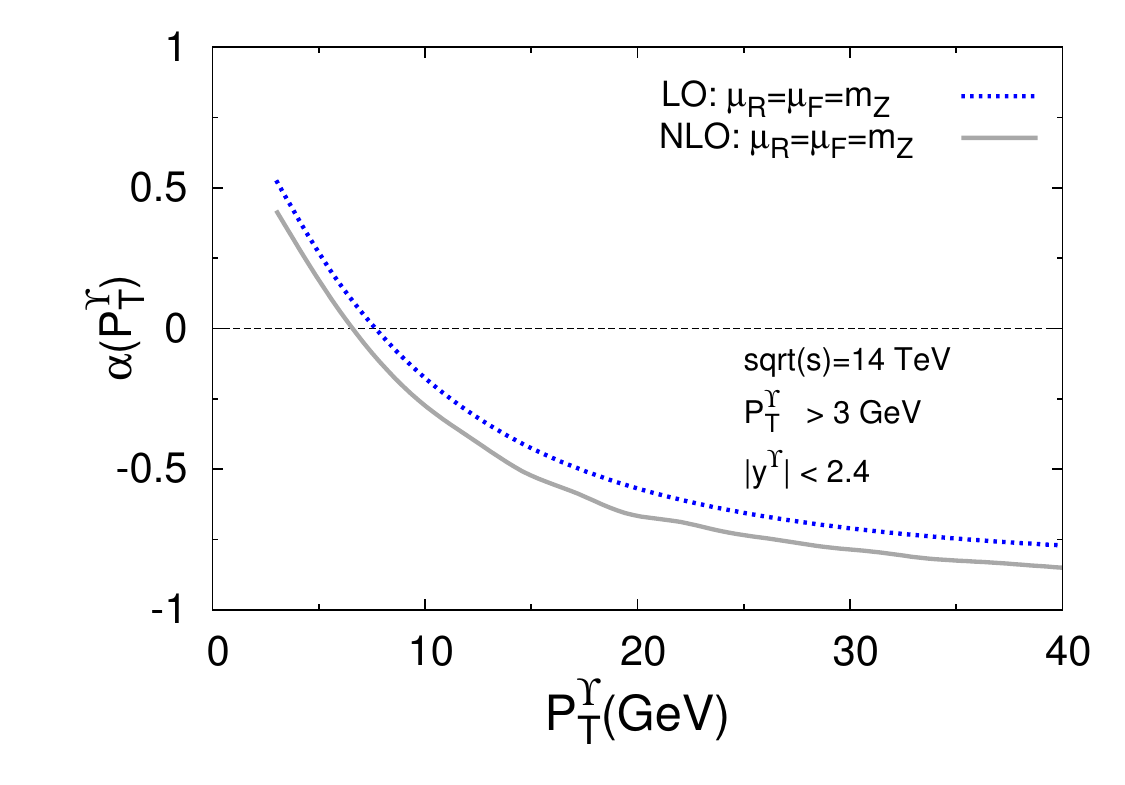}
\caption{$P_T$ dependence of the polarisation (or azimuthal anisotropy) in the helicity frame of the direct $\Upsilon$ 
produced with a $Z$ boson at LO and NLO  at $\sqrt{s}=14$ TeV.}
\label{fig:pol-upsilon}
\end{figure}

\section{Conclusions}\label{sec:conclusion}

In conclusion, we have studied the effects of the QCD corrections to the production of direct 
$J/\psi$ and $\Upsilon$ via colour-singlet transitions in association with a $Z$ boson at the LHC. 
We have found, contrary to an earlier study~\cite{Mao:2011kf}, that the NLO QCD corrections 
are consistent with the expectations, namely increasing for increasing $P_T$ and small at low $P_T$. 
We expect that a few hundred $J/\psi+Z$ events
could be detected at the LHC at 14 TeV with 100 fb$^{-1}$ of data. At 8 TeV with 20 fb$^{-1}$, there may be 
just enough events to derive a cross section and not only an upper bound on its value. Interestingly, the CSM yield
expected for direct $\Upsilon+Z$ is of the same order of magnitude than that of direct $J/\psi+Z$ at 14 TeV, if not larger.

We have studied the scale sensitivity of the $J/\psi+Z$ cross section at LO and NLO. At low $P_T$, it is smaller 
when QCD corrections are  taken into account. On the contrary, at large $P_T$, \ie~when $P_T\,\gsim\, m_Z/2$,  the 
dominant contributions are of the kind of
 $gg\to {\cal Q}+  Z + \hbox{jet}$. These involve an additional power of $\alpha_s$ and 
the sensitivity on the renormalisation scale is larger. Along the same lines of
arguments, one expects a further increase of the CSM cross section at large $P_T$ 
when the leading-$P_T$ contribution contained in the NNLO
topologies are accounted for. That being said, the presence of the $Z$ boson mass renders the CSM prediction more precise
at low $P_T$ compared to the inclusive case, for which the leading-$P_T$ contributions at NLO and NNLO dominate at lower $P_T$.

We have also found that the yield polarisation is not altered by the QCD corrections. From this observation, 
we have concluded that when a $Z$ boson is emitted by one of the heavy quarks forming the quarkonium, both 
the $J/\psi$ and the $\Upsilon$ are longitudinally polarised at LO and NLO, 
thus independently of the off-shellness and of the transverse momentum of the gluons producing the heavy-quark pair. 
This is at odds with the cases of inclusive $\cal Q$ production and $\cal Q +\gamma$ production, and this 
motivates further theoretical and experimental investigations.


\section*{Acknowledgements}

This work is supported in part by the France-China Particle Physics Laboratory, by the P2I network and by the National Natural Science Foundation of China (Nos. 10979056 and 11005137).



\begin{thebibliography}{99}



\bibitem{Kramer:2001hh}
  M.~Kramer,
  Prog.\ Part.\ Nucl.\ Phys.\  {\bf 47}, 141 (2001)
  [hep-ph/0106120].

\bibitem{Brambilla:2004wf}
  N.~Brambilla {\it et al.}, CERN Yellow Report 2005-005, 
hep-ph/0412158


\bibitem{Lansberg:2006dh}
  J.~P.~Lansberg,
  Int.\ J.\ Mod.\ Phys.\  A {\bf 21}, 3857 (2006)
  [hep-ph/0602091].



%
\bibitem{Campbell:2007ws}
  J.~Campbell, F.~Maltoni and F.~Tramontano,
  Phys.\ Rev.\ Lett.\  {\bf 98}, 252002 (2007)
  [hep-ph/0703113].



\bibitem{Artoisenet:2007xi}
  P.~Artoisenet, J.~P.~Lansberg and F.~Maltoni,
  Phys.\ Lett.\  B {\bf 653}, 60 (2007)
  [hep-ph/0703129].


\bibitem{Gong:2008sn}
  B.~Gong and J.~X.~Wang,
  Phys.\ Rev.\ Lett.\  {\bf 100} (2008) 232001.
  [arXiv:0802.3727 [hep-ph]];
%
\bibitem{Gong:2008hk}
  B.~Gong and J.~X.~Wang,
  Phys.\ Rev.\  D {\bf 78} (2008) 074011.
  [arXiv:0805.2469 [hep-ph]].


\bibitem{Artoisenet:2008fc}
  P.~Artoisenet, J.~Campbell, J.~P.~Lansberg, F.~Maltoni and F.~Tramontano,
  Phys.\ Rev.\ Lett.\  {\bf 101} (2008) 152001.
  [0806.3282 [hep-ph]].


\bibitem{CSM_hadron}
C-H. Chang,
{Nucl. Phys. } B {\bf 172} (1980) 425; 
R. Baier and R. R\"uckl,
{Phys. Lett. } B {\bf 102} (1981) 364; 
R. Baier and R. R\"uckl,
{Z. Phys. } C {\bf 19} (1983) 251.

\bibitem{Lansberg:2008gk}
  J.~P.~Lansberg,
  Eur.\ Phys.\ J.\ C {\bf 61} (2009) 693
  [arXiv:0811.4005 [hep-ph]].


\bibitem{Brambilla:2010cs}
  N.~Brambilla,  \etal~ 
  Eur.\ Phys.\ J.\ C {\bf 71} (2011) 1534
  [arXiv:1010.5827 [hep-ph]].


\bibitem{ConesadelValle:2011fw}
  Z.~Conesa del Valle, G.~Corcella, F.~Fleuret, E.~G.~Ferreiro, V.~Kartvelishvili, B.~Kopeliovich, J.~P.~Lansberg and C.~Lourenco {\it et al.},
  Nucl.\ Phys.\ (PS)  {\bf 214} (2011) 3
  [arXiv:1105.4545 [hep-ph]].



\bibitem{Li:2008ym}
  R.~Li and J.~X.~Wang,
  Phys.\ Lett.\  B {\bf 672} (2009) 51.
  [arXiv:0811.0963 [hep-ph]].


\bibitem{Lansberg:2009db}
  J.~P.~Lansberg,
  Phys.\ Lett.\  B {\bf 679} (2009) 340.
  [arXiv:0901.4777 [hep-ph]].



\bibitem{Leibovich:1996pa}
  A.~K.~Leibovich,
  Phys.\ Rev.\ D {\bf 56} (1997) 4412
  [hep-ph/9610381].

\bibitem{Kim:1994bm}
  C.~S.~Kim and E.~Mirkes,
  Phys.\ Rev.\ D {\bf 51} (1995) 3340
  [hep-ph/9407318].



\bibitem{Brodsky:2009cf}
  S.~J.~Brodsky and J.~P.~Lansberg,
  Phys.\ Rev.\ D {\bf 81} 051502(R) (2010).
 [arXiv:0908.0754 [hep-ph]].

\bibitem{Lansberg:2010cn}
  J.~P.~Lansberg,
  PoS ICHEP {\bf 2010} (2010) 206
  [arXiv:1012.2815 [hep-ph]].

\bibitem{Lansberg:2012ta}
  J.~P.~Lansberg,
  to appear in Nucl.\ Phys.\ A. [arXiv:1209.0331 [hep-ph]].

\bibitem{Adare:2006kf}
  A.~Adare {\it et al.}, 
  Phys.\ Rev.\ Lett.\  {\bf 98} 232002 (2007) .
  [arXiv:hep-ex/0611020].

\bibitem{Abelev:2010am}
  B.~I.~Abelev {\it et al.}  [STAR Coll.],
  Phys.\ Rev.\ D {\bf 82} (2010) 012004
  [arXiv:1001.2745 [nucl-ex]].

\bibitem{Acosta:2001gv}
  D.~E.~Acosta {\it et al.}  [CDF Coll.],
  Phys.\ Rev.\ Lett.\  {\bf 88}, 161802 (2002).

\bibitem{Abazov:2005yc}
  V.~M.~Abazov {\it et al.}  [D0 Coll.],
  Phys.\ Rev.\ Lett.\  {\bf 94}, 232001 (2005)
  [Erratum-ibid.\  {\bf 100}, 049902 (2008)]
  [hep-ex/0502030].

\bibitem{Khachatryan:2010zg}
  V.~Khachatryan {\it et al.}  [CMS Coll.],
  Phys.\ Rev.\ D {\bf 83} (2011) 112004
  [arXiv:1012.5545 [hep-ex]].


\bibitem{Aad:2011xv}
  G.~Aad {\it et al.}  [ATLAS Coll.],
  Phys.\ Lett.\ B {\bf 705} (2011) 9
  [arXiv:1106.5325 [hep-ex]].



\bibitem{Aaij:2012ve}
  R.~Aaij, {\it et al.}  [The LHCb Coll.],
  Eur.\ Phys.\ J.\ C {\bf 72} (2012) 2025
  [arXiv:1202.6579 [hep-ex]].

\bibitem{Aaij:2011jh}
  R.~Aaij {\it et al.}  [LHCb Coll.],
  Eur.\ Phys.\ J.\ C {\bf 71} (2011) 1645
  [arXiv:1103.0423 [hep-ex]].
\bibitem{Aamodt:2011gj}
  K.~Aamodt {\it et al.}  [ALICE Coll.],
  Phys.\ Lett.\ B {\bf 704} (2011) 442
  [arXiv:1105.0380 [hep-ex]].


\bibitem{Cooper:2004qe}
  F.~Cooper, M.~X.~Liu and G.~C.~Nayak,
  Phys.\ Rev.\ Lett.\  {\bf 93} (2004) 171801
  [hep-ph/0402219].



\bibitem{ee}
Z.~G.~He, Y.~Fan and K.~T.~Chao,
  Phys.\ Rev.\  D {\bf 81} (2010) 054036.
  [arXiv:0910.3636 [hep-ph]].
Y.~J.~Zhang, Y.~Q.~Ma, K.~Wang and K.~T.~Chao,
 Phys.\ Rev.\  D {\bf 81} (2010) 034015.
  [arXiv:0911.2166 [hep-ph]].
Y.~Q.~Ma, Y.~J.~Zhang and K.~T.~Chao,
Phys.\ Rev.\ Lett.\ {\bf 102} (2009) 162002;
[arXiv:0812.5106 [hep-ph]].
B.~Gong and J.~X.~Wang,
Phys.\ Rev.\ Lett.\ {\bf 102} (2009) 162003.
[arXiv:0901.0117 [hep-ph]].

\bibitem{Maltoni:2006yp}
  F.~Maltoni, \etal\ 
  Phys.\ Lett.\ B {\bf 638} (2006) 202
  [hep-ph/0601203].

\bibitem{Mao:2011kf}
  S.~Mao, M.~Wen-Gan, L.~Gang, Z.~Ren-You and G.~Lei,
  JHEP {\bf 1102} (2011) 071
  [arXiv:1102.0398 [hep-ph]].







\bibitem{Gong:2009ng}
  B.~Gong and J.~-X.~Wang,
  Phys.\ Rev.\ D {\bf 80} (2009) 054015
  [arXiv:0904.1103 [hep-ph]].

\bibitem{Wang:2004du}
  J.~-X.~Wang,
  Nucl.\ Instrum.\ Meth.\ A {\bf 534} (2004) 241
  [hep-ph/0407058].

\bibitem{Duplancic:2003tv}
  G.~Duplancic and B.~Nizic,
  Eur.\ Phys.\ J.\ C {\bf 35} (2004) 105
  [hep-ph/0303184].


\bibitem{Harris:2001sx}
  B.~W.~Harris and J.~F.~Owens,
  Phys.\ Rev.\ D {\bf 65} (2002) 094032
  [hep-ph/0102128].



\bibitem{Altarelli:1979ub}
  G.~Altarelli, R.~K.~Ellis and G.~Martinelli,
  Nucl.\ Phys.\ B {\bf 157} (1979) 461.



\bibitem{Artoisenet:2007qm}
  P.~Artoisenet, F.~Maltoni and T.~Stelzer,
  JHEP {\bf 0802} (2008) 102.
  [0712.2770 [hep-ph]].

\bibitem{Madonia}
{\small MADONIA}  can be used online (model ``Quarkonium production in SM'') at {\tt \footnotesize http://madgraph.hep.uiuc.edu}.



\bibitem{Campbell:2003dd}
  J.~M.~Campbell, R.~K.~Ellis, F.~Maltoni and S.~Willenbrock,
  Phys.\ Rev.\ D {\bf 69} (2004) 074021
  [hep-ph/0312024].

\bibitem{lhcb:2012dz}
  LHCb Collaboration,
  JHEP {\bf 1206} (2012) 141
  [arXiv:1205.0975 [hep-ex]].


\bibitem{Beneke:1998re}
  M.~Beneke, M.~Kramer and M.~Vanttinen,
  Phys.\ Rev.\ D {\bf 57} (1998) 4258
  [hep-ph/9709376].

\bibitem{Faccioli:2010kd}
  P.~Faccioli, C.~Lourenco, J.~Seixas and H.~K.~Wohri,
  Eur.\ Phys.\ J.\ C {\bf 69} (2010) 657
  [arXiv:1006.2738 [hep-ph]].

\bibitem{Faccioli:2012nv}
  P.~Faccioli,
  Mod.\ Phys.\ Lett.\ A {\bf 27} (2012) 1230022
  [arXiv:1207.2050 [hep-ph]].


\bibitem{Acosta:2003mu}
  D.~Acosta {\it et al.}  [CDF Collaboration],
  Phys.\ Rev.\ Lett.\  {\bf 90} (2003) 221803.

\bibitem{private} M. Kruse, A. Limosani, C. Zhou, private communications.


\bibitem{Cho:1995ce}
  P.~L.~Cho and A.~K.~Leibovich,
  Phys.\ Rev.\ D {\bf 53} (1996) 6203
  [hep-ph/9511315].

\bibitem{Braaten:1998th}
  E.~Braaten, J.~Lee and S.~Fleming,
  Phys.\ Rev.\ D {\bf 60} (1999) 091501
  [hep-ph/9812505].

\bibitem{Gong:2010bk}
  B.~Gong, J.~-X.~Wang and H.~-F.~Zhang,
  Phys.\ Rev.\ D {\bf 83} (2011) 114021
  [arXiv:1009.3839 [hep-ph]].

\bibitem{Khoze:2004eu}
  V.~A.~Khoze, A.~D.~Martin, M.~G.~Ryskin and W.~J.~Stirling,
  Eur.\ Phys.\ J.\ C {\bf 39} (2005) 163
  [hep-ph/0410020].

\expandafter\ifx\csname natexlab\endcsname\relax\def\natexlab#1{#1}\fi
\expandafter\ifx\csname bibnamefont\endcsname\relax
  \def\bibnamefont#1{#1}\fi
\expandafter\ifx\csname bibfnamefont\endcsname\relax
  \def\bibfnamefont#1{#1}\fi
\expandafter\ifx\csname citenamefont\endcsname\relax
  \def\citenamefont#1{#1}\fi
\expandafter\ifx\csname url\endcsname\relax
  \def\url#1{\texttt{#1}}\fi
\expandafter\ifx\csname urlprefix\endcsname\relax\def\urlprefix{URL }\fi
\providecommand{\bibinfo}[2]{#2}
\providecommand{\eprint}[2][]{\url{#2}}

\end{thebibliography}
\end{document}